\documentclass[prd,onecolumn]{revtex4}
\usepackage{dcolumn}
\usepackage{array}
\usepackage{multirow}
\usepackage{booktabs}
\usepackage{graphicx}
\usepackage{amssymb}
\usepackage{bm}
\usepackage{hyperref}
\usepackage{epstopdf}
\usepackage{color}
\usepackage{mathrsfs}
\usepackage{amsmath,amssymb,amsthm}
\begin{document}
\title{First comprehensive constraints on the Finslerian models using cosmological observations}
\author{Deng Wang}
\email{Cstar@mail.nankai.edu.cn}
\affiliation{Theoretical Physics Division, Chern Institute of Mathematics, Nankai University,
Tianjin 300071, China}
\author{Xin-He Meng }
\email{xhm@nankai.edu.cn}
\affiliation{Department of Physics, Nankai University, Tianjin 300071, China}
\begin{abstract}
We perform the first comprehensive constraints on two Finslerian models, i.e., the simplest Finslerian $\Lambda$CDM (F$\Lambda$) and non-flat F$\Lambda$ models using the Type Ia supernovae, baryonic acoustic oscillations, cosmic microwave background and lensing observations. Using the most stringent constraints we can provide, we find that the constrained typical parameters of both Finslerian models are all consistent with zero at the $2\sigma$ confidence level (C.L.). This means that both Finslerian models just deviate very slightly from the $\Lambda$CDM one, which is also verified by using two geometrical diagnostics to distinguish three different models from each other. We also find that a spatially flat universe is still preferred in the framework of Finsler geometry, that our measured values of the spectral index $n_s$ of primordial power spectrum in both Fisnlerian models exclude the scale invariance at more than $8\sigma$ C.L., and that the current $H_0$ tension can be relieved from $3.4\sigma$ to $2.9\sigma$ and $2.8\sigma$ in the F$\Lambda$ and non-flat F$\Lambda$ models, respectively.

\end{abstract}
\maketitle
\section{Introduction}
Modern cosmology has entered a new era driven by `` big data ''. In the past about two decades, with gradually mounting data, a number of cosmological observations such as Type Ia supernovae (SNIa) \cite{1,2}, baryonic acoustic oscillations (BAO) \cite{3}, cosmic microwave background (CMB) anisotropies \cite{4,5}, etc., have strongly indicated that the universe is undergoing a phase of accelerated expansion. The simplest scenario to explain the intriguing phenomena is a combination of the cosmological constant and cold dark matter component ($\Lambda$CDM model) \cite{6}, which is still in remarkably good agreement with almost all cosmological data more than ten years after the observational discovery of the accelerated expansion rate of the universe. Nonetheless, this very successful model still faces two unsolved and attractive puzzles \cite{7}: (i) Why is the value of the cosmological constant unexpectedly small with respect to any physically meaningful scale, except the current horizon scale ? (ii) Why this value is not only small, but also surprisingly close to another unrelated physical quantity, the present matter density ? Since we are still unclear about the nature of dark sector of the universe, to alleviate or even solve these puzzles, a great deal of alternative cosmological models based on different physical origins are proposed. In general, they can be divided into two main classes, the dark energy model \cite{8,9,10,11,12,13,14,15,16,17,18,19,20,21,22,23,wm,24}, which introduces a new fluid or field in the universe, and the extended theory of gravity \cite{25,26,27,28,29,30,31,32,33,34,35}, which modifies the standard lagrangian of Einstein's gravity. Very interestingly, one can also address the above puzzles by exploring the connection between gravitation and new geometry. Finsler geometry \cite{24,25,26}, which includes Riemann geometry as its special case, is a good candidate to understand the current cosmological puzzles. This new geometry keeps the elegant properties of Riemann geometry, i.e., the corresponding isometric group is a Lie group on a Finslerian manifold, while it admits less Killing vectors than a Riemannian spacetime does. Generally, there are $n(n-1)/2+1$ independent Killing vectors in a $n$ dimensional non-Riemannian Finslerian spacetime at most. Taking the simplest possible asymmetrical generalization of Riemannian metric into account, G. Randers \cite{36} proposed the so-called Randers space, a subclass of Finslerian space. In the framework of Randers space, a generalized Friedmann-Robertson-Walker (FRW) cosmological model based on Finsler geometry has been studied \cite{37}, and a modified dispersion relation of free particles has also been discussed \cite{38}.

The gravity in a Finslerian space were studied for a long time \cite{39,40,41,42}. The gravitational field equations (GFEs) derived from a Riemannian osculating metric were presented in Ref. \cite{43}. For such a metric, the FRW-like cosmological scenario and the anisotropies of the universe were also investigated \cite{37,44}. However, their GFEs are derived without satisfying the Bianchi identity and the general covariance principle of Einstein's gravity. Attractively, the authors in Refs. \cite{45,46,47} have solved these problems and derived the corresponding Friedmann-like equations in Ref. \cite{45} by constructing a Randers-Finsler space of approximate Berwald type, which is just an extension of a Riemannian space. Following this theoretical line, we are motivated by placing constraints on the new Finslerian cosmological model using the current cosmological observations.

This study is organized in the following manner. In Section 2, we introduce briefly the Finslerian models to be constrained. In Section 3, we describe the methodology and data we use in this analysis, and exhibit the corresponding constraining results. In Section 4, we distinguish this model from the $\Lambda$CDM model using two geometrical diagnostics. The concluding remarks are presented in the final section.

\section{Finslerian models}
In this study, we consider two Finslerian models, i.e., the simplest Finslerian $\Lambda$CDM (F$\Lambda$) model and its one-parameter extension, non-flat F$\Lambda$ model. In a Finsler-Berwald FRW universe, combining the time-component Friedmann equation and the acceleration equation \cite{45,46,47}, the continuity equation can be expressed as
\begin{equation}
\dot{\rho}(\frac{3\alpha}{4}+\frac{\beta}{12}+1)+\dot{p}(-\frac{3\alpha}{4}+\frac{\beta}{4})=-\frac{\dot{a}}{a}[\rho(\frac{3\alpha}{2}+\frac{\beta}{6}+2)+p(-\frac{3\alpha}{2}+\frac{\beta}{2})+(\rho+3p)(1+\frac{\beta}{3})], \label{1}
\end{equation}
where $a$, $\alpha$ and $\beta$ are the scale factor, effective time-component and space-component parameters, respectively. Note that here \textit{`` effective ''} means that the physical quantities are derived based on the non-Riemannian Berwald space. Substituting the equation of state (EoS) $p_i=\omega_i\rho_i$ of each independent component $i$ (where the constant $\omega_i$ corresponds to the non-relativistic matter, dark energy and effective curvature, respectively) into Eq. (\ref{1}), one can obtain the effective energy density as follows
\begin{equation}
\rho_i\propto a^{-\frac{36(1+\omega_i)+18(1-\omega_i)\alpha+6(1+3\omega_i)\beta}{12+9(1-\omega_i)\alpha+(1+3\omega_i)\beta}}. \label{2}
\end{equation}
Subsequently, using the time-component Friedmann equation and Eq. (\ref{2}) the square of the effective dimensionless Hubble parameter $E_1(a)$ of the F$\Lambda$ model can be written as
\begin{equation}
E_1^2(a)=\frac{9\alpha+\beta+12}{4(\alpha+1)(\beta+3)}\Omega_{m}a^{-\frac{6(3\alpha+\beta+6)}{9\alpha+\beta+12}}+\frac{9\alpha-\beta+6}{2(\alpha+1)(\beta+3)}\Omega_{de}            a^{-\frac{6(3\alpha-\beta)}{9\alpha-\beta+6}},         \label{3}
\end{equation}
where $\Omega_{m}$ and $\Omega_{de}$ denote the effective matter and dark energy density parameters today, respectively. Note that this model reduces to the $\Lambda$CDM model when $\alpha=\beta=0$. Since we are of interest in investigating the evolution of the late universe, we ignore the contribution from the radiation component. Furthermore, considering the spatial curvature in the Finslerian scenario, the square of the effective dimensionless Hubble parameter $E_2(a)$ of the non-flat F$\Lambda$ model is expressed as
\begin{equation}
E_2^2(a)=\frac{3}{3+\beta}\Omega_{k}a^{-2}+ \frac{9\alpha+\beta+12}{4(\alpha+1)(\beta+3)}\Omega_{m}a^{-\frac{6(3\alpha+\beta+6)}{9\alpha+\beta+12}}+\frac{9\alpha-\beta+6}{2(\alpha+1)(\beta+3)}\Omega_{de}            a^{-\frac{6(3\alpha-\beta)}{9\alpha-\beta+6}},         \label{4}
\end{equation}
where $\Omega_{k}$ is the effective curvature density parameter today. One can also find that the contribution from curvature based on Finsler geometry depends only on the space-component parameter $\beta$. We use units $8\pi G=c=1$ throughout this work.
\begin{figure}
\centering
\includegraphics[scale=0.4]{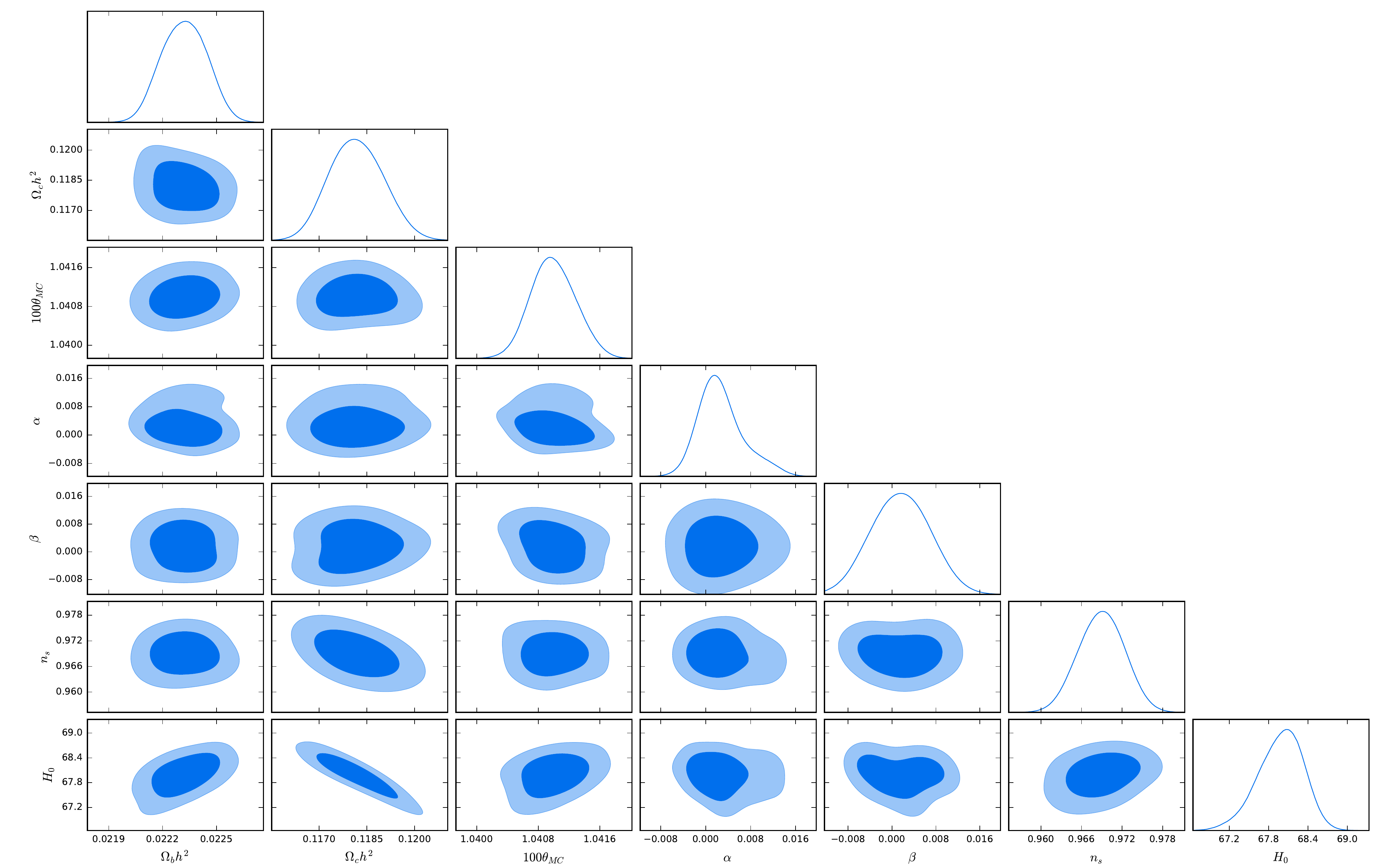}
\caption{The one-dimensional marginalized probability distribution on the individual parameter and $2$-dimensional contours of the F$\Lambda$ model by using SBC data sets.}\label{f1}
\end{figure}
\begin{figure}
\centering
\includegraphics[scale=0.4]{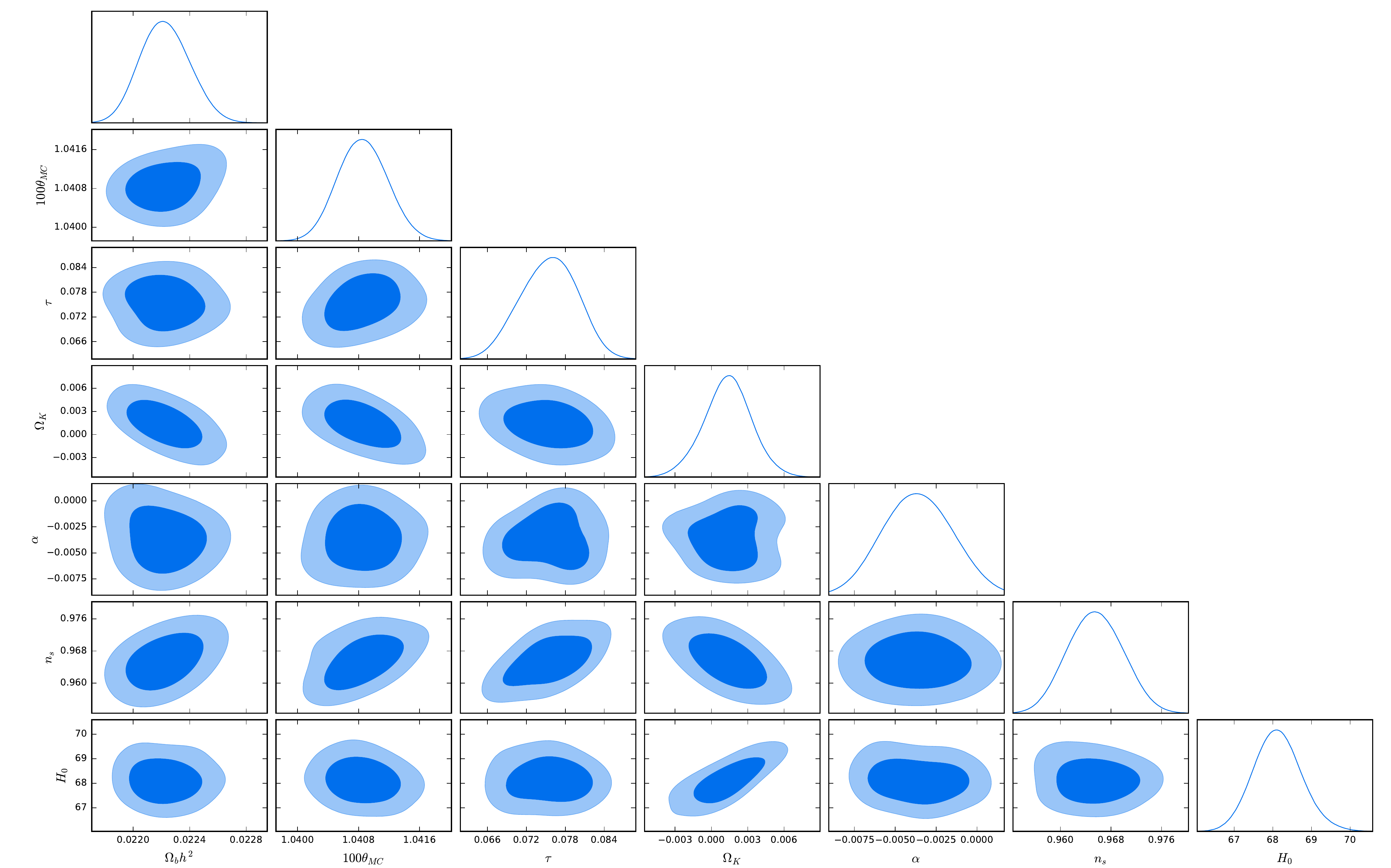}
\caption{The one-dimensional marginalized probability distribution on the individual parameter and $2$-dimensional contours of the non-flat F$\Lambda$ model  using SBC data sets.}\label{f2}
\end{figure}
\begin{figure}
\centering
\includegraphics[scale=0.4]{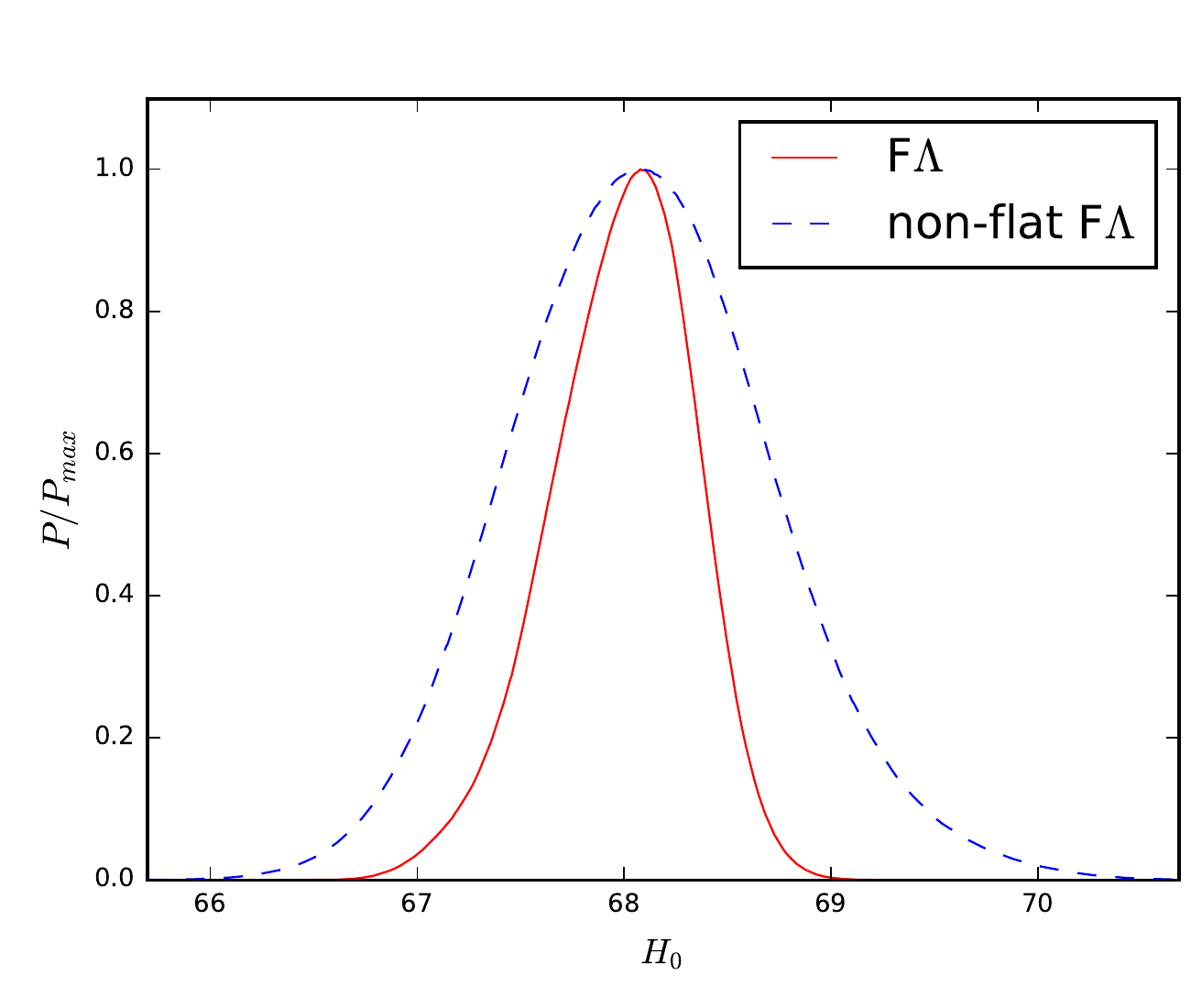}
\caption{The one-dimensional posterior probability distributions of $H_0$ values derived from the F$\Lambda$ (solid) and non-flat F$\Lambda$ (dashed) models using SBC data sets, respectively.}\label{f3}
\end{figure}

\section{Constraints}
We employ the Markov Chain Monte Carlo (MCMC) method to place constraints on the above two Finslerian models by using the current cosmological observations. More specifically, we modify carefully the publicly MCMC code CosmoMC \cite{48,49} and Boltzmann code CAMB \cite{50} to infer the posterior probability distributions of parameters, and analyze the correlated MCMC samples using the publicly code GetDist \cite{51}. Meanwhile, we choose flat priors to the model parameters and marginalize the foreground nuisance parameters provided by Planck. For the F$\Lambda$ model, we consider the following 8-dimensional parameter space
\begin{equation}
\mathbf{P_1}=\{\Omega_bh^2, \quad \Omega_ch^2, \quad 100\theta_{MC}, \quad \tau, \quad  \alpha, \quad  \beta, \quad  \mathrm{ln}(10^{10}A_s), \quad  n_s\},  \label{5}
\end{equation}
where $\Omega_bh^2$ and $\Omega_ch^2$ denote, respectively, the present-day baryon and CDM densities, $100\theta_{MC}$ is the approximation to $100 \times$  angular size of sound horizon at the redshift of last scattering $z_*$, $\tau$ represents the Thomson scattering optical depth due to reionization, $\alpha$ and $\beta$ are two typical model parameters of F$\Lambda$ model, $\mathrm{ln}(10^{10}A_s)$ and $n_s$ denote the amplitude and the spectral index of the primordial scalar perturbation power spectrum at the pivot scale $K_0=0.05$ Mpc$^{-1}$, respectively. Here $h\equiv H_0/100$ and $H_0$ is the Hubble constant. Similarly, the 9-dimensional parameter space for the non-flat F$\Lambda$ model is expressed as
\begin{equation}
\mathbf{P_2}=\{\Omega_bh^2, \quad \Omega_ch^2, \quad 100\theta_{MC}, \quad \Omega_k,  \quad \tau, \quad  \alpha, \quad  \beta, \quad  \mathrm{ln}(10^{10}A_s), \quad  n_s\},  \label{6}
\end{equation}
where $\Omega_{k}$ denotes the effective curvature density today. Subsequently, we exhibit the cosmic probes used in this analysis, including the SNIa, BAO, CMB and Planck lensing data.

$\bullet$ The SNIa data: Since the absolute magnitudes of all the SNIa are considered to be the same based on the nearly same explosion mass, the SNIa is regarded as a standard candle to explore the background evolution of the universe in theory. We use the largest `` Joint Light-curve Analysis '' (JLA) sample containing 740 SNIa data points, which covers the redshift range $z \in [0.01, 1.3]$ \cite{52}. The JLA compilation consists of 118 low-$z$ SNe in the range $z \in [0, 0.1]$ from \cite{53,54,55,56,57,58}, 374 SNe in the range $z \in [0.3, 0.4]$ from the Sloan Digital Sky Survey (SDSS) SNe search \cite{59},  239 SNe in the range $z \in [0.1, 1.1]$ from the Supernova Legacy Survey (SNLS) project \cite{60}, and 9 high-$z$ SNe in the range $z \in [0.8, 1.3]$ from the Hubble Space Telescope (HST) \cite{61}.

$\bullet$ The BAO data: To break the parameter degeneracy efficiently, we take four BAO measurements consisting of the 6dFGS (six-degree-field galaxy survey) sample at effective redshift $z_{eff}=0.106$ \cite{62}, the SDSS MGS (main galaxy sample) sample at $z_{eff}=0.15$ \cite{63}, and the LOWZ at $z_{eff}=0.32$ and CMASS $z_{eff}=0.57$ samples of the SDSS-III BOSS (Baryon Oscillation Spectroscopic Survey) DR12 sample \cite{64}.

$\bullet$ The CMB data: We also use the Planck 2015 temperature and polarization data in our numerical analysis \cite{65}, including the likelihoods of temperature (TT) at $30\leqslant \ell\leqslant 2500$, the cross correlation of temperature and polarization (TE), the polarization (EE) power spectra and the Planck low-$\ell$ temperature and polarization likelihood at $2\leqslant \ell\leqslant 29$.

$\bullet$ The lensing data: We utilize the Planck lensing data \cite{66}, which gives the most powerful measurement to date with a 2.5$\%$ constraint on the amplitude of the lensing potential power spectrum (or alternatively, a 40$\sigma$ detection of lensing effects).

\begin{table}[h!]
\caption{The prior ranges of different model parameters used in the Bayesian analysis.}
\label{t1}
\begin{tabular}{ll}

\hline
\hline
Parameter            &Prior                                         \\
\hline
$\Omega_bh^2$              &$[0.005, 0.1]$                   \\
$\Omega_ch^2$              &$[0.001, 0.99]$                   \\
$100\theta_{MC}$            &$[0.5, 10]$                      \\
$\tau$                    &$[0.01, 0.8]$                           \\
$\Omega_k$                &$[-0.05, 0.05]$                   \\
$\alpha$                    &$[-3, 3]$                         \\
$\beta$                    &$[-3, 3]$                                \\
$\mathrm{ln}[10^{10}A_s]$        &$[2, 4]$         \\
$n_s$                   &$[0.8, 1.2]$                          \\
$H_0$                   &$[20, 100]$                           \\
\hline
\hline
\end{tabular}
\end{table}

\begin{table}[h!]
\renewcommand\arraystretch{1.3}

\caption{The $1\sigma$ ($68\%$) uncertainties of different parameters in the F$\Lambda$ and non-flat F$\Lambda$ models using the data combinations S, SB and SBC, respectively.}
\label{t2}
\begin{tabular} { l c c c c c c}

\hline
\hline
Model & & F$\Lambda$ && & non-flat F$\Lambda$ \\
\hline
Data & S & SB &  SBC & S & SB &  SBC    \\
\hline
{$\Omega_b h^2   $} & $0.0232^{+0.0016}_{-0.0014}$ & $0.02239\pm 0.00084        $ & $0.02232\pm 0.00012        $ & $0.0235^{+0.0023}_{-0.0026}$  & $0.02183\pm 0.00052        $ & $0.02222\pm 0.00016        $                                                        \\

{$\Omega_c h^2   $} & $0.1169^{+0.0027}_{-0.0037}$ & $0.1193^{+0.0043}_{-0.0071}$ & $0.11817\pm 0.00081        $ & $0.110^{+0.011}_{-0.014}   $ & $0.1204^{+0.0036}_{-0.0040}$  & $0.1193^{+0.0016}_{-0.0013}$                                                          \\

{$100\theta_{MC} $} & $1.04016\pm 0.00097        $ & $1.0416^{+0.0047}_{-0.0065}$ & $1.04100^{+0.00028}_{-0.00031}$  & $1.04273^{+0.00095}_{-0.0024}$ & $1.0401^{+0.0017}_{-0.0012}$  & $1.04085\pm 0.00031        $                                                      \\

{$\tau           $} &$0.130\pm 0.029             $ & $0.153^{+0.082}_{-0.031}   $ & $0.0810^{+0.0052}_{-0.0072}$ & $0.148\pm 0.046                   $ & $0.076^{+0.027}_{-0.052}   $ & $0.0755^{+0.0047}_{-0.0042}$
                                                        \\
{$\Omega_K       $}  & ---  & --- & ---  & $-0.011^{+0.022}_{-0.019}  $  & $0.0028^{+0.0043}_{-0.0048}$  & $0.0013\pm 0.0019          $                                                                                                           \\

{$\alpha         $} & $0.0131\pm 0.0058          $ & $0.0013^{+0.0012}_{-0.0091}$ & $0.0027^{+0.0029}_{-0.0047}$  & $-0.001^{+0.062}_{-0.022}  $ & $-0.0024^{+0.0057}_{-0.0065}$  & $-0.0036\pm 0.0019         $                                                   \\

{$\beta          $} & $0.0000\pm 0.0032          $ & $-0.0011^{+0.0089}_{-0.0061}$ & $0.0016\pm 0.0048          $  & $-0.002^{+0.016}_{-0.026}  $ & $0.0022\pm 0.0038          $  & $-0.0051^{+0.0076}_{-0.0069}$                                                    \\

{${\rm{ln}}(10^{10} A_s)$} & $3.104^{+0.015}_{-0.010}   $ & $3.102^{+0.013}_{-0.031}   $ & $3.094^{+0.011}_{-0.023}   $  & $3.088^{+0.016}_{-0.010}   $ & $3.0994^{+0.0084}_{-0.0064}$  & $3.0848^{+0.0071}_{-0.0061}$                                             \\

{$n_s            $} & $0.949^{+0.017}_{-0.020}   $ & $0.969^{+0.011}_{-0.017}   $ & $0.9690\pm 0.0033          $ & $0.971\pm 0.067            $ & $0.9468\pm 0.0088          $   & $0.9655\pm 0.0042          $                                                     \\
\hline
$H_0              $ & $68.9^{+2.1}_{-1.9}        $ & $67.95^{+0.87}_{-1.0}      $ & $67.99^{+0.38}_{-0.30}     $  & $67.4\pm 4.9               $ & $67.84\pm 0.81             $ & $68.12\pm 0.61             $                                                      \\

\hline
\hline
\end{tabular}
\end{table}

In what follows, for simplicity, we denote the data combinations SNIa, CMB+lensing, SNIa+BAO and SNIa+BAO+CMB+lensing as S, C, SB and SBC, respectively. Specifically, we perform constraints on the F$\Lambda$ and non-flat F$\Lambda$ models by using S, SB and SBC, respectively. The prior ranges of different model parameters considered in this analysis are shown in Tab. \ref{t1}.

By implementing a Bayesian analysis, the constraining results of the above two models using different data combinations are exhibited in Tab. \ref{t2}. Utilizing the joint constraints from SBC, the corresponding one-dimensional marginalized probability distribution on the individual parameter and $2$-dimensional contours for the F$\Lambda$ and non-flat F$\Lambda$ models are shown in Figs. \ref{f1}-\ref{f2}, respectively. For the F$\Lambda$ model, we find that the values of typical model parameters $\alpha$ and $\beta$ are well consistent with zero at the $1\sigma$ C.L. and prefer slightly the positive best-fit values being very close to zero when using the data combination SBC. Being different from the F$\Lambda$ model, for the non-flat F$\Lambda$ one, the value of $\beta$ is well consistent with zero at the $1\sigma$ C.L., but the value of $\alpha$ is not. Attractively, adding a curvature parameter into the F$\Lambda$ model, the values of $\alpha$ and $\beta$ prefer slightly negative best-fit values being very close to zero when using SBC. We also find that the value of the spatial curvature $\Omega_k=0.0013\pm 0.0019$ from SBC in the non-flat F$\Lambda$ model is consistent with zero at the $1\sigma$ C.L., which means that a flat universe is still preferred in the framework of Finsler geometry. It is worth noting that, using SBC, our measured values of the spectral index $n_s$ of the primordial scalar perturbation power spectrum in the above two Fisnlerian scenarios both rule out the scale invariance at more than $8\sigma$ C.L., which is very compatible with the predictions from Planck temperature and polarization data and Planck lensing data \cite{65,66} (see Tab. \ref{t2}).

Recently, the directly local measurement $H_0=73.24\pm1.74$ km s$^{-1}$ Mpc$^{-1}$ from Riess et al. 2016 (hereafter R16) \cite{67} using the improved SNIa calibration techniques exhibits a strong tension with the indirectly global measurement $H_0=66.93\pm0.62$ km s$^{-1}$ Mpc$^{-1}$ derived by Planck collaboration  (hereafter P15) \cite{68} under the assumption of the base six-parameter $\Lambda$CDM model at the $3.4\sigma$ level. We are also of much interest in addressing this issue using the Finslerian settings. Using SBC data sets, the one-dimensional posterior probability distributions of $H_0$ values derived from the F$\Lambda$ and non-flat F$\Lambda$ models are shown in Fig. \ref{f3}. Interestingly, we find that, using SBC, the current $H_0$ tension can be alleviated from $3.4\sigma$ to $2.9\sigma$ and $2.8\sigma$ in the F$\Lambda$ and non-flat F$\Lambda$ models, respectively. By opening an extra parameter $\Omega_k$, we also find, as expected, a mild relaxation on $H_0$ limits at the $1\sigma$ C.L. in the non-flat F$\Lambda$ model.
\begin{figure}
\centering
\includegraphics[scale=0.5]{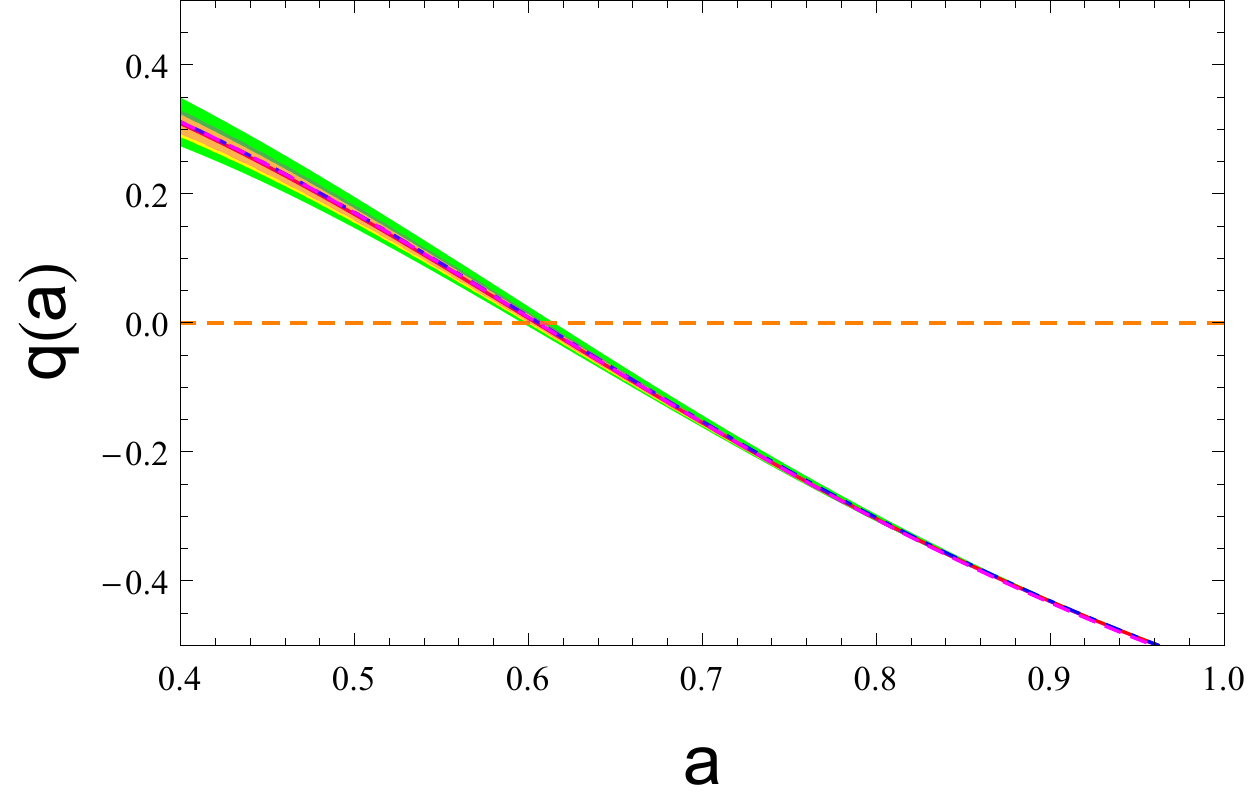}
\includegraphics[scale=0.5]{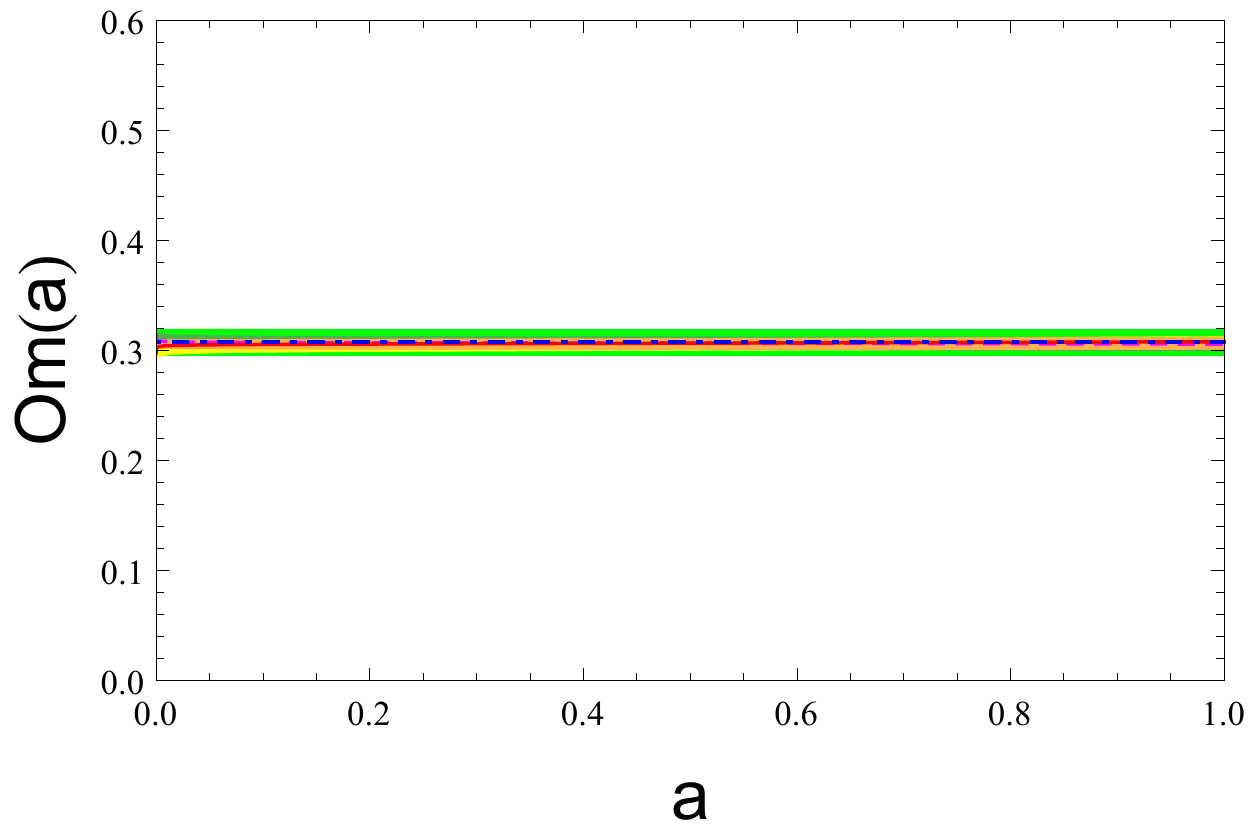}
\caption{The left and right panels correspond to the relations between the scale factor $a$ and the acceleration parameter $q(a)$ and $Om(a)$ diagnostic, respectively. The red (solid), blue (dash-dotted) and magenta (short-dashed) lines correspond to the best-fit F$\Lambda$, non-flat F$\Lambda$ and $\Lambda$CDM models, respectively. The shaded yellow, purple and green regions represent the $1\sigma$ regions of the F$\Lambda$, non-flat F$\Lambda$ and $\Lambda$CDM models, respectively. The horizontal orange (long-dashed) line corresponds to the zero-acceleration universe.}\label{f4}
\end{figure}
\section{Diagnostics}
In this section, we employ two simple geometrical diagnostics, i.e., the deceleration parameter and $Om(a)$ diagnostic \cite{69} to distinguish the F$\Lambda$ and non-flat F$\Lambda$ models from the standard cosmological model and one from the other. The deceleration parameter for a given dark energy model can be expressed as
\begin{equation}
q(a)=-\frac{a}{E(a)}\frac{dE(a)}{da}-1,  \label{7}
\end{equation}
where $E(a)$ of the above two Finslerian models can be found in Eqs. (\ref{3}-\ref{4}).

The $Om(a)$ diagnostic is an very useful geometrical method to distinguish different cosmological models from each other, which is written as
\begin{equation}
Om(a)=\frac{E^2(a)-1}{a^{-3}-1}.  \label{8}
\end{equation}
It is easy to verify that, for a flat $\Lambda$CDM model, the $Om(a)$ diagnostic is fixed to be $\Omega_{m}$, namely $Om(a)=\Omega_{m}$ (see also Ref. \cite{69}).

Utilizing $\Omega_{m}=0.308\pm 0.012$ measured by Planck collaboration \cite{65} for the $\Lambda$CDM model and error propagations of the constrained model parameters $\alpha$, $\beta$, $\Omega_m$ and $\Omega_k$ using SBC data sets, we present the diagnostic results in Fig. \ref{f4}. We find that the evolutional behaviors of two Finslerian models are very close to that of the $\Lambda$CDM one, and consequently, they cannot be distinguished from the $\Lambda$CDM model at the $1\sigma$ C.L.. By simple calculations, we also obtain the deceleration-acceleration redshift $z_{t1}=0.6558^{+0.0130}_{-0.0103}$ for the F$\Lambda$ model and $z_{t2}=0.6517^{+0.0121}_{-0.0102}$ for the non-flat F$\Lambda$ model at the $1\sigma$ C.L., respectively.

\section{Concluding remarks}
To understand accurately the current cosmic acceleration from a new geometrical perspective, we have performed the first constraints on the simplest Finslerian model, the F$\Lambda$ model and its one-parameter extension, the non-flat F$\Lambda$ model by using the current cosmological observations. Utilizing the most stringent constraints SBC we can provide, we find that: For the F$\Lambda$ model, the values of typical model parameters $\alpha$ and $\beta$ are well consistent with zero at the $1\sigma$ confidence level and prefer slightly the positive best-fit values being very close to zero; For the non-flat F$\Lambda$ one, nonetheless, the values of $\alpha$ and $\beta$ prefer slightly negative best-fit values being very close to zero and the value of $\beta$ is well consistent with zero at the $1\sigma$ C.L., but the value of $\alpha$ is not; A spatially flat universe is still preferred in the framework of Finsler geometry; Our measured values of the spectral index $n_s$ of primordial power spectrum in both Fisnlerian scenarios rule out the scale invariance at more than $8\sigma$ C.L.;
The current $H_0$ tension can be relived from $3.4\sigma$ to $2.9\sigma$ and $2.8\sigma$ in the F$\Lambda$ and non-flat F$\Lambda$ models, respectively.

Using two popular geometrical diagnostics, we find that both Finslerian models cannot be distinguished from the $\Lambda$CDM model during the evolutional process of the universe. It is noteworthy that we obtain this conclusion just at the background evolution level without considering the perturbation effects. Furthermore, in this analysis, we just test primarily the abilities of two Finslerian models in explaining the current cosmological phenomena without consider other interesting extensions. In addition, more data except SBC can also be applied in constraining the Finslerian models. In total, our investigations on using the Finsler geometry to understand the evolution of the universe and reconcile the cosmological tensions among various probes are just in the beginning stage. The remaining issues will be addressed carefully in a forthcoming study \cite{70}.

\section*{ACKNOWLEDGMENTS}
Deng Wang warmly thanks Jing-Ling Chen and F. Canfora for very useful communications. Xin-He Meng appreciates S. D. Odintsov and B. Ratra for helpful discussions on cosmology. This study is partly supported by the National Science Foundation of China.

\begin{appendix}
\section{u}
\end{appendix}

\end{document}